\documentclass[14pt]{article}
\usepackage[utf8]{inputenc}
\usepackage{tgtermes}
\usepackage{lipsum}
\usepackage{cite}
\usepackage{authblk}
\usepackage[symbol]{footmisc}
\usepackage{graphicx}
\usepackage{subfig}
\usepackage[margin=1in]{geometry}
\usepackage{amsmath}
\usepackage{amssymb}
\usepackage[nottoc]{tocbibind}
\usepackage{amssymb}
\usepackage{physics}
\usepackage{cancel}
\usepackage{yfonts}
\usepackage{tikz}
\usepackage{mwe}

\usepackage{geometry}
\usepackage{hyperref}
\definecolor{darkslateblue}{rgb}{0.28, 0.24, 0.55}

\hypersetup{colorlinks,linkcolor={darkslateblue}, citecolor={magenta}}  

\makeatletter
\def\@fnsymbol#1{\ensuremath{\ifcase#1\or \ddagger\or \dagger\or
   \mathsection\or \mathparagraph\or \|\or **\or \ddagger\ddagger
   \or \dagger\dagger \else\@ctrerr\fi}}

\title{Minimal length corrections to magnetic birefringence in vacuum}
\author{Ribhu Paul \footnote{~~ribhupaul.rp@gmail.com / psrp2452@iacs.res.in}}
\affil{{\small{{{School of Physical Sciences\\
{Indian Association for the Cultivation of Science, Kolkata - 700032, India}}}}}}
\date{}

\begin{document}

\maketitle


\begin{abstract}
    We consider the modification to \textit{Kruglov}'s non-linear electrodynamics in the presence of minimal length ($l_p$), by considering a specific formulation of generalized uncertainty principle (GUP). The presence of minimal length has been motivated by several candidate theories of quantum-gravity (both perturbative and non-perturbative). We show that, such minimal length modifications could lead to momentum dependent bounds on zero-point electric field strength, for the aforementioned non-linear model of electrodynamics. We further show that, the existence of minimal length modifies the measure of vacuum birefringence $\Delta n$ in a constant and uniformly strong, external magnetic field. We therefore suggest, the presence of a frictional mechanism of fundamental origin, acting in the parallel direction of polarization of photons (in the presence of strong magnetic field), as an additional effect of quantized background. 
    
\end{abstract}


\section{Introduction}

A consistent theoretical framework for quantum-gravity, features domain of extreme scales (where quantum-gravitational effects are predominant) that are inevitable, leading to profound epistemological and experimental challenges for physicists \cite{kiefer2013conceptual}. The current empirical limitations and theoretical freedom, have led to some candidate theories (LQG, string theory, quantum geometrodynamics and others) that are fundamentally different \cite{ziaeepour2022comparing}. As suggested and motivated in string theory, non-commutative geometries and LQG \cite{hossenfelder2006note,ashtekar2004background,douglas2001noncommutative,girelli2005deformed,gross1988string,thiemann2003lectures,yoneya1989interpretation,perez2003spin,konishi1990minimum, veneziano1986stringy}, existence of a minimal length $l_p$, is a physical requirement in the regime of quantum gravity. One possible analysis of phenomenological aspects (in presence of minimal length), can be done by deforming \textit{Heisenberg}'s uncertainty principle (GUP) in the \textit{perturbative} regime \cite{bosso2022minimal}, which eventually leads to quasi-locality \cite{kempf1995hilbert} and modified \textit{Heisenberg (Kempf)} algebra, as a function of deformation parameters $\beta$ and $\beta'$\cite{quesne2006lorentz}. In particular, by considering such algebraic modifications, a minimal length formulation of electrodynamics in vacuum (under a specific deformation constraint, $\beta'=2\beta$), has been studied by \textit{Moayedi} et al.\cite{moayedi2013formulation} which includes the modified \textit{Maxwell}'s equations (in free space) and dispersion relation. \\[4pt]
After 1930s, a series of theoretical investigations carried out by \textit{W. Heisenberg} and \textit{H. Euler} \cite{heisenberg2006consequences} suggested that, under the influence of a sufficiently strong magnetic perturbation, change in the linear behavior of quantum vacuum occurs which results in vacuum birefringence (analogous to \textit{Cotton-Mouton} effect in liquids \& gases where the birefringence is linear). Several non-linear models have been proposed for studying the strong magnetic birefringence effect in vacuum \cite{kruglov2015nonlinear}. From analysis, it is found that, the difference between the refractive indices of parallel and transverse polarization directions i.e. $\Delta n=n_{\parallel}-n_{\perp}$, is related to the strength of the external magnetic field  producing birefringence. Such vacuum birefringence predictions in formidable magnetic fields \cite{Hattori_2013}, could be a possible observable in the domain of astrophysics of isolated neutron stars \cite{mignani2017evidence}, where \textit{Schwinger} limit of magnetic field is often exceeded. However, it is questionable whether quantum-gravitational effects still preserve similar behavior for $\Delta n$ in the same range. In this regard, we thus explore the possible corrections to vacuum birefringence.\\[4pt] In the following sections, we firstly review the covariant deformed algebra (section- \ref{section-2}). Then, we motivate an effective $Lagrangian$ for minimally corrected electromagnetism (section- \ref{section-3}). Lastly, we consider a generic non-linear model of electrodynamics (section- \ref{section-4}) as suggested by \textit{Kruglov} \cite{kruglov2015nonlinear}, where vacuum birefringence is theoretically investigated in the presence of a constant, strong magnetic perturbation, after incorporating the ingredients of minimal length from GUP. The zero point field strength is also studied in this context.\\[4pt] In this article, we therefore conclude that, on imposing minimal length corrections to \textit{Kruglov}'s formulation of non-linear electrodynamics\cite{kruglov2015nonlinear}, a correction is obtained for $\Delta n$. Moreover, we obtain that this correction is dependent on the strength of the photon momentum. Hence, we argue that, the functional behavior of $\Delta n$ with a greater value might be a signature of quantum-gravitational, resistive force, possibly arising from the discrete background. 
\\[2pt]
We shall use the metric convention \textit{diag}($1,-1,-1,-1$) and the \textit{Levi-Civita} convention $\epsilon_{0123}=+1$ all throughout. Greek indices run over space-time components whereas, Latin indices run over space components alone. In the final part of the computation, we have considered $c=1, \mu_0=1 ~\&~ \epsilon_0=1$  for simplicity, which has been mentioned where needed. \\


\section{Covariant deformed algebra in minimal length background \textit{(Kempf-Quesne-Tkachuk} algebra)}\label{section-2}

A formal discussion on modification to \textit{Heisenberg's} uncertainty principle in the presence of a minimal length, and the resulting deformed algebra was done by \textit{Kempf} \cite{kempf1995hilbert}. Later, \textit{Quesne} and \textit{Tkachuk} obtained a covariant generalization of the same. The generalized uncertainty principle in 1+D-dimensional quantized space-time\cite{quesne2006lorentz}, which is often motivated from perturbative string theory\cite{gross1988string} and LQG is the following\footnote[3]{The following specific form of generalized uncertainty principle can also be motivated from duality principle of the zero-point length of space-time \cite{mondal2020duality}.}:

\begin{equation}\label{equation-1}
\begin{aligned}
      \Delta X^i \Delta P^i \ge \frac{\hbar}{2}\Bigl(1 +& f^{i}_{D}(\beta, \beta')\Bigr)~,\\
      \text{where,~~~}
f^{i}_{D}(\beta, \beta')=\beta'\left[(\Delta P)^2+\langle P^i \rangle^2\right]-\beta \Bigl\{\langle& (P^0)^2\rangle-\sum_{j=1}^D~\left[  (\Delta P)^2+\langle P^i \rangle^2\right]   \Bigr\}~~,\\ \text{and $i=1,2,......,D$}~~,\\
\end{aligned}
\end{equation}
for isotropic uncertainties --- $\Delta P^i = \Delta P;~ \forall i$. The above inequality is retained upto first order in deformation parameters --- $
\beta$ and $\beta'$ which are related to minimal length (from eq.:\ref{equation-3}) and have the dimension --- (momentum)$^{-2}$. A deformed \textit{Lorentz} covariant algebra $\mathcal{K}_{cov}^D(\beta, \beta')$ is thus obtained as follows \cite{quesne2006lorentz}:
\begin{equation}\label{equation-2}
    \begin{aligned}
          \relax [X^{\mu},P^{\nu}]&=-i\hbar \left[(1-\beta P_{\rho}P^{\rho}){\eta}^{\mu \nu} - \beta' P^{\mu}P^{\nu}\right]~~,\\
          [X^{\mu},X^{\nu}]&= i\hbar\left[ \frac{2\beta-\beta'}{1-\beta P_{\rho}P^{\rho} }(P^{\mu}X^{\nu}-P^{\nu}X^{\mu})+\mathcal{O}(\beta^2,\beta'^2,\beta \beta')\right]~~,\\
          [P^{\mu},P^{\nu}]&=0~~.
    \end{aligned}
\end{equation}
It is obvious that the commutativity of $X^{\mu}$ and $X^{\nu}$ i.e. $[X^{\mu},X^{\nu}]=0$, is maintained in the first order for deformation constraint $\beta'=2\beta$. We shall follow this prescription all throughout.\\[2pt]
Now, from the generalized uncertainty principle expressed as a formal expansion in $l_p^2$ \cite{moayedi2013formulation}:
\begin{equation}\label{equation-3}
    \Delta X^i \Delta P^i \ge \frac{\hbar}{2}\Bigl(1+a_1^i(D)\left(\frac{l_p}{\hbar}\right)^2(\Delta P)^2 + \mathcal{O}\bigl(l_p^4,(\Delta P)^4\bigr)\Bigr)~~\text{~~where, $a^i_1$ are functions of $D$,}
\end{equation}
and (eq.-  \ref{equation-1}), the deformation parameter $\beta$ can be defined as a quadratic function of $l_p$ in the leading order i.e. $\beta=\mathcal{O}(l_p^2)$.\\[2pt]
From (eq.- \ref{equation-2}) and the deformation prescription, a possible representation of $X^{\mu}$ and $P^{\nu}$ is obtained as follows\cite{moayedi2013formulation}:
\begin{equation}\label{equation-4}
    \begin{aligned}
    X^{\mu}&=x^{\mu}~~ \text{and}\\
    P^{\nu}&=(1-\beta p_{\rho}p^{\rho})p^{\nu}~~,
    \end{aligned}
\end{equation}
where $x^{\mu}$ and $p^{\mu}=i\hbar \displaystyle{\frac{\partial}{\partial x_{\mu}}\equiv i\hbar \partial^{\mu}}$, are the usual \textit{Heisenberg} representation of canonical pair of operators. 
It is to be mentioned that, in the position representation, $X^{\mu}$ and $P^{\nu}$ are given as:
\begin{equation}\label{equation-5}
    X^{\mu}=x^{\mu}\text{~~~and~~~}P^{\nu}\equiv i\hbar~\textfrak{D}^{\nu}=i\hbar(1+\beta \hbar^2 \Box_D)\partial^{\nu}~~, 
\end{equation}
where $\Box_D\equiv \partial_{\rho}\partial^{\rho}$ is the 1+D-dimensional $d'Alembertian$.


\section{Minimal length formulation of electrodynamics in vacuum by effective Lagrangian}\label{section-3}
In order to study the effects on electromagnetic waves, moving in quantized 1+3-dimensional background in the absence of charges, we need to formulate a suitable description of electromagnetism in the presence of minimum observable (physical) length. The \textit{Lagrangian} for free electromagnetic field in the absence of minimal length is given by:
\begin{equation*}
    \mathcal{L}_{EM}^{(0)}=-\frac{1}{4\mu_0}F^{\alpha \beta}F_{\alpha \beta}~~,
\end{equation*}
where $F_{\alpha \beta}=\partial_{\alpha} A_{\beta}-\partial_{\beta} A_{\alpha}$ is the \textit{Faraday} tensor and $A^{\alpha}\equiv (\phi/c, \mathbf{A})$ is the 4-potential. It is to be noted that the entire treatment retains the classicality of the fields.\\
Now, using (\ref{equation-5}) and by straight-forward substitution as done by \textit{Moayedi} et al.\cite{moayedi2013formulation}:
\begin{equation}\label{equation-6}
    \begin{aligned}
   x^{\mu}&\rightarrow X^{\mu}~~\\
   \partial^{\nu}&\rightarrow \textfrak{D}^{\nu}~~,\\
    \end{aligned}~~
\end{equation}
it can be shown that:
\begin{equation}\label{equation-6a}
    \begin{aligned}
       \mathcal{L}_{EM}^{(0)}\rightarrow \mathcal{L}_{EM}^{(0)}-\frac{1}{4\mu_0} a^2 F_{\alpha \beta}\Box F^{\alpha \beta}\equiv\mathcal{L}_{EM}^{(a^2)}~~,
    \end{aligned}
\end{equation}
where, $\Box \equiv \Box_3$ and $a\equiv \hbar \sqrt{2\beta}$ is \textit{Podolsky}'s characteristic length. A theoretical cut-off results in the existence of a maximum value for $a$. Precisely, the bound $|a|<< 4.72 \cross 10^{-16}$ cm $\equiv a_m$ \cite{frenkel1999self,accioly2010limits} sets an upper bound to the deformation parameter $\beta$. The above form of the \textit{Lagrangian} is related to the formulation of generalized  electrodynamics by $Podolsky$, capable of describing \textit{Land\'e-Thomas} theory as a special case \cite{podolsky1942generalized}.\\

The minimal formulation can yet be motivated in a different way, which we shall establish in the following part. It is to be noted that the implications from GUP are very much physical and is representation independent. Hence, its consistency is justified only in the perturbative regime \cite{bosso2022minimal}. The above substitution (from \ref{equation-6}) renders a representation dependent form of the original \textit{Faraday} tensor that often curtails its physical significance. We therefore, consider a general effective form of \textit{Lagrangian} $\mathcal{L}^{(a^2)}_{eff}$ under minimal length setting without modifying the form of $Faraday$ tensor, keeping a note that the leading order correction to the \textit{Lagrangian} should be of $\mathcal{O}(\beta)$ i.e. $\mathcal{O}(a^2)$ or $\mathcal{O}(l_p^2)$. We also hope to retain the perturbative correction for the same in the range $\displaystyle{\beta < \frac{a_m^2}{2\hbar^2}}$. 
\begin{equation}
    \begin{aligned}
    \text{Let us consider~~~~~~~~}
    \mathfrak{F}^{\alpha \beta}&=F^{\alpha \beta} + i\left(\frac{a}{a_{m}}\right) \Tilde{F}^{\alpha \beta}~~,\\
    \end{aligned}
\end{equation}
where $F^{\alpha \beta}$ and $\Tilde{F}^{\alpha \beta}$ are real valued and $\Tilde{F}^{\alpha \beta}\equiv \partial^{\alpha}\Tilde{A}^{\beta}-\partial^{\beta}\Tilde{A}^{\alpha}$. \\
In other words, a simple substitution of
\begin{equation*}
    A^{\alpha}\rightarrow \mathfrak{A}^{\alpha} = A^{\alpha}+i\left(\frac{a}{a_m}\right)\Tilde{A}^{\alpha}~~,
\end{equation*}
serves the same mathematical requirement but carries over the representation dependence to $\mathfrak{A}^{\alpha}$ (see eq.- \ref{equation-10}). It is worth mentioning that the electric and magnetic field components are related to the real part of $\mathfrak{A}^{\alpha}$ and the imaginary part of $\mathfrak{A}^{\alpha}$ accounts for the minimal correction. Also, $\Tilde{A}$ must be determined in terms of $A$, which essentially follows from the principles of perturbative approaches,i.e. one could consider $\mathfrak{A}_{\alpha}^{(0)}\equiv A_{\alpha}$ and $\mathfrak{A}_{\alpha}^{(1)}\equiv \Tilde{A}_{\alpha}$, then determine  $\mathfrak{A}_{\alpha}^{(1)}$ from $\mathfrak{A}_{\beta}^{(0)}$ using perturbative methods~. Hence, it is reasonable to assume the auxilliary part $\Tilde{A}_{\alpha}=\Tilde{A}_{\alpha}(A_{\beta})$.

Now, let us consider the following effective \textit{Lagrangian} that yields first order correction to the bilinear invariant $F_{\alpha \beta}F^{\alpha \beta}$: 

\begin{equation*}
-\frac{1}{4\mu_0}\mathfrak{F}^{*}_{\alpha \beta} \mathfrak{F}^{\alpha \beta}=-\frac{1}{4\mu_0}F_{\alpha \beta}F^{\alpha \beta}-\frac{a^2}{4\mu_0a_m^2}\Tilde{F}_{\alpha \beta}\Tilde{F}^{\alpha \beta}\equiv \mathcal{L}^{(a^2)}_{eff}~~
\end{equation*} 
after retaining upto $\mathcal{O}(a^2)$.
One can show that the invariance of $\mathcal{L}^{(a^2)}_{eff}$ under~ $\mathfrak{A}^{\alpha} \rightarrow e^{i\phi}~\mathfrak{A}^{\alpha}$ (for $\phi$ restricted to real axis) reflects the mixing of \textit{Faraday} tensor with its minimal correction thereby treating both on an equal physical footing. \\
We therefore formulate the effective action functional:

\begin{equation}
    \begin{aligned}
    \mathcal{S}_{eff}^{(a^2)}\equiv \int~d^4x~\mathcal{L}_{eff}^{(a^2)}=- \int~d^4x~\frac{1}{4\mu_0}\mathfrak{F}^{*}_{\alpha \beta} \mathfrak{F}^{\alpha \beta}~~.
    \end{aligned}
\end{equation}
By setting $\displaystyle{\frac{\delta \mathcal{S}_{eff}^{(a^2)}}{\delta A_{\mu}}}=0$, we find the following equations of motion:
\begin{equation}\label{equation-9}
    \begin{aligned}
    \partial_{\mu}\Bigl( F^{\mu \nu} + \frac{a^2}{a_m^2} ~\frac{d \Tilde{A}_{\lambda}}{d A_{\nu}}~ \Tilde{F}^{\mu \lambda} \Bigr)=0
    \end{aligned}~~.
\end{equation}
Now, considering the particular choice of representation as in (eq.- \ref{equation-6}) is equivalent to imposing the  \textit{on-shell} criteria:
\begin{equation}\label{equation-10}
    \Tilde{F}^{\mu \nu}= a^2_{m}~\frac{d A_{\rho}}{d \Tilde{A}_{\nu} }~\Box F^{\mu \rho}~~,
\end{equation}
In principle, one could consider a different \textit{on-shell} criteria that would correspond to another representation of $\mathcal{K}_{cov}^3(\beta,2\beta)$. In this particular case, we observe that for $\Tilde{F}^{\mu \nu}$ and hence $S_{eff}^{(a^2)}$ to be well-defined, we must have:  $$\left|\frac{d \Tilde{A}_{\lambda}}{d {A}_{\nu} }\right| \ne 0 ~~\forall~\lambda~\text{and}~ \nu~~.$$
This implies that $\Tilde{A}_{\lambda}$ should be strictly increasing or decreasing functions of $A_{\nu}$. Now, using (eq.- \ref{equation-9}) and (eq.- \ref{equation-10}), we arrive at the modified \textit{Maxwell}'s equations \cite{moayedi2013formulation} :
\begin{equation}\label{equation-11}
    \Bigl(\partial_{\mu} F^{\mu \nu} + {a^2}~ \Box~ \partial_{\mu} F^{\mu \nu} \Bigr)=0
\end{equation}


To summarize, we obtain the modified \textit{Maxwell}'s equations by constructing a general effective action of $\mathcal{O}(a^2)$. Further, we consider imposing a specific condition on $\mathcal{I}m(\mathfrak{A}^{\alpha})$ to be equivalent to choosing a particular representation. It is to be noted that, the above formulation preserves the manifestation of \textit{Faraday} tensor as it would be in the absence of minimal length and the modifications are carried over to the imaginary part of $\mathfrak{A}^{\alpha}$ by introducing the \textit{on-shell} criteria separately. It must also be mentioned that the choice of a specific representation corresponds to a particular curve on the complex plane $\mathfrak{A}^{\alpha}$ (see eqn.- \ref{equation-10}). \\[2pt]
Thus, we obtain the modified wave equation that are directly obtainable from the minimally corrected \textit{Maxwell}'s equations (eq.- \ref{equation-11})  \cite{moayedi2013formulation} :
\begin{equation}
    \begin{aligned}
    \Box~ \mathbf{E} + a^2 \Box \Box~ \mathbf{E}&=0\\
    \Box~ \mathbf{B} + a^2 \Box \Box~ \mathbf{B}&=0~~.\\
    \end{aligned}
\end{equation}
It is straightforward to obtain the solution for $\mathbf{B}$ from above, after plugging in the following ansatz: $$\mathbf{B}=\mathcal{B}~e^{-ip_{\nu}x^{\nu}/\hbar}~~.$$
We therefore, obtain the plane wave solution in minimal setting (similar arguments are valid for $\mathbf{E}$):
\begin{equation}\label{equation-14}
    \begin{aligned}
    \mathbf{B}^{(a)}(\mathbf{x},t)&=\mathcal{B}~e^{i\mathbf{p}\cdot \mathbf{x}/\hbar}~e^{-i\Bar{E}^{(a)}_{\mathbf{p}}t/\hbar}\\
    &=\mathbf{B}^{(0)}(\mathbf{x},t)~e^{i(E^{(0)}_{\mathbf{p}}-\Bar{E}^{(a)}_{\mathbf{p}})t/\hbar}\equiv \mathbf{B}^{(0)}(\mathbf{x},t)~e^{i\mathfrak{\omega^{(a)}_{\mathbf{p}}}t}~~,
    \end{aligned}
\end{equation}
where, $$p_{\mu}p^{\mu}\equiv p^2 = \frac{\hbar^2}{a^2}\implies \Bar{E} _{\mathbf{p}}^{(a)} =\pm \abs{\mathbf{p}}c~\sqrt{1+\frac{\hbar^2}{a^2 \abs{\mathbf{p}}^2}}~~ $$
is the effective (modified) dispersion relation for massive (effectively) case \cite{moayedi2013formulation} ,
\begin{equation*}
    \mathbf{B}^{(0)}(\mathbf{x},t)=\mathcal{B}~e^{i\mathbf{p}\cdot \mathbf{x}/\hbar}~ e^{-iE^{(0)}_{\mathbf{p}}t/\hbar} = \mathcal{B}~e^{i(\mathbf{p}\cdot \mathbf{x}~\pm~ \abs{\mathbf{p}}ct)/\hbar}~~
\end{equation*}
are the unmodified free field solutions that satisfy:
\begin{equation}
    \Box~ \mathbf{B}^{(0)} = 0~~,
\end{equation}
and $\displaystyle{{\abs{\hbar\omega^{(a)}_{\mathbf{p}}}}}=\displaystyle{\abs{\mathbf{p}}c\left(1-\sqrt{1+\frac{\hbar^2}{a^2 \abs{\mathbf{p}}^2}}\right)}$.  \\
If we consider the expansion of the above expression, in the region of complex plane  $\Gamma$, of the dimensionless parameter $\displaystyle{\Lambda_{a}\equiv \frac{a}{a_m-a}}$ where $\displaystyle{\frac{\hbar}{{\abs{\mathbf{p}}}}<< a}$, it can be shown that the residue of $\displaystyle{\omega^{(\Lambda_a)}_{\mathbf{p}}}$ is related to $a_m$ (upper bound of \textit{Podolsky}'s characteristic length) in terms of the following contour integral defined on the boundary of the region  $\partial \Gamma$:
\begin{equation}
    \frac{\hbar}{2\pi\abs{\mathbf{p}}c} \oint_{\partial \Gamma}~d\Lambda_a~i\omega_{{\mathbf{p}}}(\Lambda_a) = \sum_{n=1}^{\infty}~~{(-1)^n}\frac{2n}{1-2n}~^{2n}\mathbf{C}_n\left(\frac{\hbar}{2\abs{\mathbf{p}}}\right)^{2n}{a_m}^{-2n}
\end{equation}
From the previous results (eqn.- \ref{equation-14}), we also observe that:
 \begin{equation}
     \begin{aligned}
           \Box~ \mathbf{B}^{(a)}(\mathbf{x},t)& + \left(\frac{\omega^{(a)}_{\mathbf{p}}}{c}\right)^2~ \mathbf{B}^{(a)}(\mathbf{x},t)=0\\
            \Box~ \mathbf{E}^{(a)}(\mathbf{x},t)& + \left(\frac{\omega^{(a)}_{\mathbf{p}}}{c}\right)^2~ \mathbf{E}^{(a)}(\mathbf{x},t)=0~~.
     \end{aligned}
 \end{equation}
 The above set of equations dictate the propagation of the modified magnetic and electric field solutions.
Next, we look into the modification of a proposed non-linear model \cite{kruglov2015nonlinear} in minimal length setting.


\section{Modification to \textit{Kruglov}'s non-linear electrodynamics}\label{section-4}
The post-\textit{Maxwellian} / perturbative approach to QED establishes some successful theoretical predictions that were experimentally verified later. Although, vacuum birefringence in the presence of a strong external field, is just another theoretical prediction that lacks sufficient empirical justification. The external field that provokes such quantum process cannot be generated in physical laboratories. Extremely strong external magnetic field is often generated in some astrophysical objects. However, such fields often exceed the perturbative (\textit{Schwinger}) limit, demanding a non-perturbative approach \cite{denisov2017nonperturbative}. 

Several semi-classical treatments of vacuum birefringence are available in the literature till date, which feature non-linearities in the classical electromagnetic theory arising from quantum vacuum effects. We shall now consider a hybrid non-linear model that reduces to different known models under suitable approximations.\\[2pt]

In order to study vacuum birefringence under constant external magnetic field, a non-linear model of electrodynamics with three parameters was suggested by \textit{Kruglov} \cite{kruglov2015nonlinear}.  The parameters introduced in the model are directly associated with the physical bounds on the electric field. We first express the \textit{Lagrangian} in its usual form:
\begin{equation}\label{equation-18}
    \mathcal{L}^{(0)}_{K}=-\mathcal{F}-\frac{\alpha \mathcal{F}}{2b\mathcal{F}+1}+\frac{\gamma}{2}\mathcal{G}^2~~,
\end{equation}
where the bilinear \textit{Lorentz} invariants are $\displaystyle{\mathcal{F}\equiv\frac{1}{4\mu_0}F_{\alpha \beta}F^{\alpha \beta}}=\frac{1}{2\mu_0}\abs{\mathbf{B}}^2-\frac{\epsilon_0}{2}\abs{\mathbf{E}}^2$, $\displaystyle{\mathcal{G}
\equiv\frac{1}{4\mu_0}}\Tilde{F}_{\alpha \beta}F^{\alpha \beta}=\frac{1}{c\mu_0}\mathbf{E}\cdot \mathbf{B}$ and $\Tilde{F}_{\alpha \beta}=\frac{1}{2}\epsilon_{\alpha \beta \delta \gamma}F^{\delta \gamma}$ is the \textit{Hodge} dual of $F_{\alpha \beta}$. It is to be noted that, $\alpha$ is the only dimensionless parameter in this description whereas, $b$ has the dimension of $(energy/length)^{-2}$ and $\gamma \mathcal{G}$ is dimensionless. 
In order to incorporate the modifications (see section- \ref{section-2}), we note that under $\mathcal{K}_{cov}^3(\beta, 2\beta')$
\begin{equation}
    \begin{aligned}
    \mathcal{F}&\rightarrow \mathcal{F}+\frac{a^2}{4\mu_0}F_{\alpha \beta}\Box F^{\alpha \beta}~~\text{and}\\
    \mathcal{G}&\rightarrow \mathcal{G}+\frac{a^2}{4\mu_0}\Tilde{F}_{\alpha \beta}\Box F^{\alpha \beta}~~.
    \end{aligned}
\end{equation}

Thus, we obtain the following modified \textit{Lagrangian}:
\begin{equation}\label{equation-19}
\begin{aligned}
        \mathcal{L}^{(a^2)}_{K}=-\mathcal{F}-\frac{a^2}{4\mu_0}&F_{\alpha \beta}\Box F^{\alpha \beta}-\frac{\alpha \mathcal{F}+\frac{\alpha a^2}{4\mu_0}F_{\alpha \beta}\Box F^{\alpha \beta}}{1+2b\mathcal{F}+\frac{ba^2}{2\mu_0}F_{\alpha \beta}\Box F^{\alpha \beta}}\\&+\frac{\gamma}{2}\mathcal{G}^2+\frac{\gamma a^2}{4\mu_0}\mathcal{G}\Tilde{F}_{\alpha \beta}\Box F^{\alpha \beta}+\mathcal{O}(a^4)~~.
\end{aligned}
\end{equation}
We notice that the presence of a minimal length supplements some additional non-linearity to the \textit{Lagrangian}.
Further, it can be shown that the $\gamma$ term is responsible for birefringence. It is to be noted that,
for $\alpha \rightarrow 0$ and $\gamma=0$ \cite{kruglov2015nonlinear}, the above \textit{Lagrangian} reduces to the familiar \textit{Lagrangian} in (eq.- \ref{equation-6a}).\\
If $\displaystyle{2b\mathcal{F}+\frac{ba^2}{4\mu_0}F_{\alpha \beta}F^{\alpha \beta}<<1}$, then the \textit{Lagrangian} simply becomes:
\begin{equation}
\begin{aligned}
    \mathcal{L}_{K}^{(a^2)}\approx-(1&+\alpha)\mathcal{F}+2\alpha b \mathcal{F}^2+\frac{\gamma}{2}\mathcal{G}^2\\&+\frac{a^2}{4\mu_0}\left[-(1+\alpha)F_{\alpha \beta}\Box F^{\alpha \beta}+4\alpha b \mathcal{F} F_{\alpha \beta}\Box F^{\alpha \beta}+\gamma \mathcal{G}\Tilde{F}_{\alpha \beta}\Box F^{\alpha \beta}\right] +\mathcal{O}(a^4)~~.
\end{aligned}
\end{equation}

Here, we observe the presence of non-linear $\mathcal{F}^2$ term in the \textit{Lagrangian} that is motivated by considering non-linear effects on quantum vacuum in higher (multi) dimensional \textit{Born-Infeld} electromagnetism\cite{lemos1999born}. Thus, in the weak $b$ approximation and $\gamma=0$, all the non-linear terms are expected to subside leaving the minimal length correction for $\mathcal{F}$. In order to have a standard quadratic term, we can redefine the above \textit{Lagrangian} in terms of the renormalized tensor $F_{\alpha \beta}'\equiv \sqrt{(1+\alpha)}F_{\alpha \beta}$ as:
\begin{equation*}
\begin{aligned}
      \mathcal{L}^{(a^2)}_{K~(REN)} \approx -\mathcal{F}'&+2\alpha b \mathcal{F}'^2+\frac{\gamma(1-\alpha)}{2}{\mathcal{G}'^2}\\&+\frac{a^2}{4\mu_0}\left[-F'_{\alpha \beta}\Box F'^{\alpha \beta}+4\alpha b \mathcal{F}' F'_{\alpha \beta}\Box F'^{\alpha \beta}+\gamma(1-2\alpha)\mathcal{G}'\Tilde{F}'_{\alpha \beta}\Box F'^{\alpha \beta}\right]+\mathcal{O}(a^4)~~,
\end{aligned}
\end{equation*}
for the natural assumption of $\alpha<<1$.\\[2pt]
It is straightforward to obtain the equations of motion for the modified \textit{Lagrangian} (in eq.- \ref{equation-19}):
\begin{equation}\label{equation-22}
    \begin{aligned}
          \partial_{\mu}\left[\left(F^{\mu \nu}+a^2\Box F^{\mu \nu}\right)+ \frac{\alpha(F^{\mu \nu}+a^2\Box F^{\mu \nu})}{\left(1+2b\mathcal{F}+\frac{ba^2}{2\mu_0}F_{\alpha \beta}\Box F^{\alpha \beta}\right)^2}-\gamma\left(\mathcal{G}\Tilde{F}^{\mu \nu}+a^2 \mathcal{G}~\Box\Tilde{F}^{\mu \nu}+\frac{a^2}{4\mu_0}\Tilde{F}^{\mu \nu}{F}_{\alpha \beta}\Box \Tilde{F}^{\alpha \beta}\right)\right]& \\[6pt]+\mathcal{O}(a^4)=0~~.\hspace{50pt}&
    \end{aligned}
\end{equation}

In the absence of the $\gamma$ term, the modified two parameter model \cite{kruglov2015model} has the following equations of motion:
\begin{equation}
    \partial_{\mu}\left[\left(F^{\mu \nu}+{a^2}\Box F^{\mu \nu}\right)+ \frac{\alpha(F^{\mu \nu}+{a^2}\Box F^{\mu \nu})}{\left(1+2b\mathcal{F}+\frac{ba^2}{2\mu_0}F_{\alpha \beta}\Box F^{\alpha \beta}\right)^2}\right]=0~~.
\end{equation}

The above equations are self consistent due to the non-linearity of the theory which is severe when a minimal length prescription is incorporated. In other words, after taking the derivative w.r.t. $\partial^{\mu}$, the equations can be re-expressed in the form as follows:
\begin{equation*}
\Box F^{\mu \nu} = f^{\mu \nu}\left(\alpha, b, \gamma; F^{\delta \rho}, a^2~\Box F^{\delta \rho}, \mathcal{F}, \mathcal{G} \right)~~,
\end{equation*}
where $f^{\mu \nu}$ is a tensorial form of function that can be obtained from the equations of motion discussed above. We therefore, seek an approximation that is reasonable enough for obtaining $F^{\mu \nu}$. Such approximation is motivated from the study of linear electrodynamics with minimal length as discussed before, which must also relate to the limiting behavior of $a$. We therefore make the following leading order approximation as well as the assertion:

\begin{equation}
\begin{aligned}
   a^2 \Box F^{\delta \rho}&\approx -a^2 \left(\frac{\Bar{\omega}_{\mathbf{p}}^{(a)}}{c}\right)^2 F^{\delta \rho}\\~~
   \text{where,}
    \left({\Bar{\omega}}_{\mathbf{p}}^{(a)}\right)^2&\equiv \left({{\omega}}_{\mathbf{p}}^{(a)}\right)^2-\frac{c^2}{a^2}
\end{aligned}
\end{equation}

As a consequence, we have:
\begin{equation*}
    \lim_{a\rightarrow 0} a^2 \Box F^{\delta \rho}\approx 0~~,
\end{equation*}
which is consistent theoretically as the minimal length corrections drop out in this limit and the equations of motion are similar to the ones obtained by \textit{Kruglov} \cite{kruglov2015model,kruglov2015nonlinear}.\\[2pt]

Under the above consideration, the equations of motion are devoid of any self-consistency and take the following form:

\begin{equation}
    \Box F^{\mu \nu} = f^{\mu \nu}\Bigl(\alpha, b, \gamma; F^{\delta \rho}, -\frac{a^2}{c^2}{\left(\Tilde{\omega}_{\mathbf{p}}^{(a)}\right)^2}, \mathcal{F}, \mathcal{G}\Bigr)~~.
\end{equation}
More precisely, we obtain:
\begin{equation}\label{equation-26}
    \begin{aligned}
         \partial_{\mu}\left[\Bigl(F^{\mu \nu}-a^2\left(\frac{\Tilde{\omega}_{\mathbf{p}}^{(a)}}{c}\right)^2 F^{\mu \nu}\Bigr)+ \frac{\alpha(F^{\mu \nu}-a^2 \left(\frac{\Tilde{\omega}_{\mathbf{p}}^{(a)}}{c}\right)^2  F^{\mu \nu})}{\Bigl(1+2b\mathcal{F}-2{ba^2}\left(\frac{\Tilde{\omega}_{\mathbf{p}}^{(a)}}{c}\right)^2 \mathcal{F}\Bigr)^2}-\gamma\Bigl(\mathcal{G}\Tilde{F}^{\mu \nu}-2a^2\left(\frac{\Tilde{\omega}_{\mathbf{p}}^{(a)}}{c}\right)^2  \mathcal{G}\Tilde{F}^{\mu \nu}\Bigr)\right]& \\[6pt]+~\mathcal{O}(a^4)= 0~~.\hspace{70pt}&
    \end{aligned}
\end{equation}
and for $\gamma=0$ (two-parameter model), we have the following equations of motion:

\begin{equation}\label{equation-27}
    \begin{aligned}
         \partial_{\mu}\left[\Bigl(F^{\mu \nu}-a^2\left(\frac{\Tilde{\omega}_{\mathbf{p}}^{(a)}}{c}\right)^2 F^{\mu \nu}\Bigr)+ \frac{\alpha(F^{\mu \nu}-a^2 \left(\frac{\Tilde{\omega}_{\mathbf{p}}^{(a)}}{c}\right)^2  F^{\mu \nu})}{\Bigl(1+2b\mathcal{F}-{2ba^2}\left(\frac{\Tilde{\omega}_{\mathbf{p}}^{(a)}}{c}\right)^2 \mathcal{F}\Bigr)^2}\right] \\[6pt]+~\mathcal{O}(a^4)= 0~~.
    \end{aligned}
\end{equation}

Equations - \ref{equation-26} along with the \textit{Bianchi} identity $\partial_{\mu} \Tilde{F}^{\mu \nu}=0$, correspond to the modified \textit{Kruglov-Maxwell}'s equations in the presence of minimal length.
From eqn.- \ref{equation-26}, we can therefore read out the electric induction field as follows: 

\begin{equation*}
    \mathbf{D}^{(a^2)}= \frac{\partial \mathcal{L}^{(a^2)}_{K}}{\partial \mathbf{E}}
\end{equation*}
Here, the electric induction field takes the following form:
\begin{equation*}
\begin{aligned}
     \mathbf{D}^{(a^2)}_{\gamma~~~i}&= \epsilon^{(a^2)}_{\gamma~~~ij} ~\mathbf{E}_{j}\\
     \text{where,~~~~~~}
     \epsilon^{(a^2)}_{\gamma~~~ij}&=\epsilon^{(a^2)}~\delta_{ij}+\Gamma^{(a^2)}_{\gamma}~ \mathbf{B}_i \mathbf{B}_j~~
     \text{,~~~~~~~~~~~~~~~~}\\
\end{aligned}
\end{equation*}
\begin{equation}\label{equation-28}
\begin{aligned}
    \epsilon^{(a^2)}=\left[\Bigg(1-a^2\left(\frac{\Tilde{\omega}_{\mathbf{p}}^{(a)}}{c}\right)^2\Bigg)+ \frac{\alpha \Bigg(1-a^2 \left(\frac{\Tilde{\omega}_{\mathbf{p}}^{(a)}}{c}\right)^2\Bigg)}{\Bigl(1-b\epsilon_0 \abs{\mathbf{E}}^2+\frac{b}{\mu_0}\abs{\mathbf{B}}^2+{ b\epsilon_0 a^2}\left(\frac{\Tilde{\omega}_{\mathbf{p}}^{(a)}}{c}\right)^2 \abs{\mathbf{E}}^2-\frac{ba^2}{\mu_0}\left(\frac{\Tilde{\omega}_{\mathbf{p}}^{(a)}}{c}\right)^2 \abs{\mathbf{B}}^2\Bigr)^2}\right]&\\
   \text{and,~~~~~~~~~~~~~~~~}
   \Gamma^{(a^2)}_{\gamma}=\frac{\gamma}{c\mu_0}\left[{1}-2a^2\left(\frac{\Tilde{\omega}_{\mathbf{p}}^{(a)}}{c}\right)^2\right]~~.\hspace{70pt} &
\end{aligned}
\end{equation}
For the ease of computation, we choose $c=1, \mu_0=1 ~\&~ \epsilon_0=1$, which can later be recovered from dimensional analysis.\\[2pt]
 In addition, it is also possible to define the magnetic field  $\mathbf{H}$ from the following canonical / standard relation: $$\displaystyle{\mathbf{H}^{(a^2)}= -\frac{\partial \mathcal{L}^{(a^2)}_{K}}{\partial \mathbf{B}}}~~,$$
as well as the magnetic permeability tensor from classical electromagnetism \cite{kruglov2015model} in minimal length background:
\begin{equation}
    \mu_{\gamma~~~ij}^{(a^2)}=\frac{1}{\epsilon^{(a^2)}}~\delta_{ij}-\frac{\Gamma^{(a^2)}_{\gamma}}{\epsilon^{(a^2)}\Bigl(\Gamma^{(a^2)}_{\gamma}\abs{\mathbf{E}}^2-\epsilon^{(a^2)}\Bigl)}~\mathbf{E}_i\mathbf{E}_j~~.
\end{equation}
Consequentially, we have:
\begin{equation*}
    \mathbf{H}^{(a^2)}_{\gamma~~~i}= \left(\mu_{\gamma}^{(a^2)}\right)^{-1}_{ij}~\mathbf{B}_j~~.
\end{equation*}

At this stage of analysis, it is possible to investigate the physical bound on the electric field \cite{kruglov2015model} by considering the two parameter model. So, we take the specific model (see eqn.- \ref{equation-27}) for $\gamma=0$,  where the electric permittivity tensor $\epsilon_{\gamma~~~ij}^{(a^2)}$ becomes isotropic as $\Gamma^{(a^2)}_{\gamma}$ being a linear function of $\gamma$ identically goes to zero i.e. $\Gamma_0^{(a^2)}=0$.\\
%

Further, if we consider the electrostatic approximation in the presence of a point source ($\mathbf{H}=\mathbf{B}=0$), we get the modified \textit{Gauss}' law as follows:
\begin{equation*}
    \nabla \cdot \mathbf{D}^{(a^2)}_0=q\delta(\mathbf{r})~~,
\end{equation*}
which yields:
\begin{equation*}
    \mathbf{D}^{(a^2)}_0(\mathbf{r})=\frac{q}{4\pi r^3}\mathbf{r}~~.
\end{equation*}
\begin{figure}[ht]
\centering 
\begin{minipage}[t]{6cm}
\centering
\includegraphics[scale=0.45]{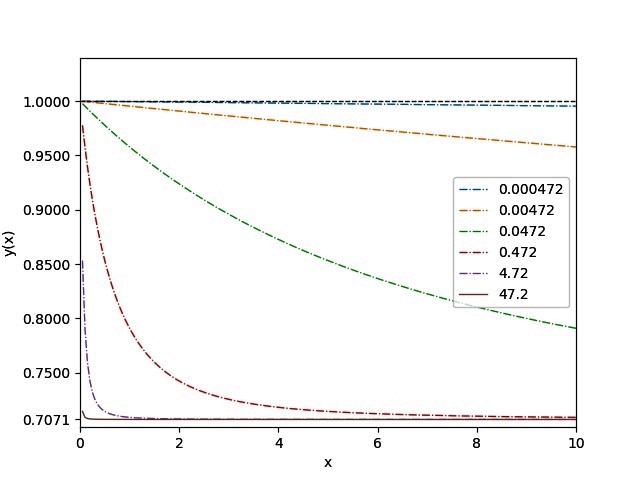}
\caption{The above schematic depicts the momentum dependent variation of field strength occurring from possible perturbative quantum gravity effects on non-linear electrodynamics. Plots for different values of $a$ in the units of $~ 10^{-17}~cm$ are constructed for comparison. The horizontal line depicts the maximal field strength in the absence of any characteristic minimum length\cite{kruglov2015model}. }
\end{minipage}
\hspace{2cm}
\begin{minipage}[t]{6cm}
\includegraphics[scale=0.45]{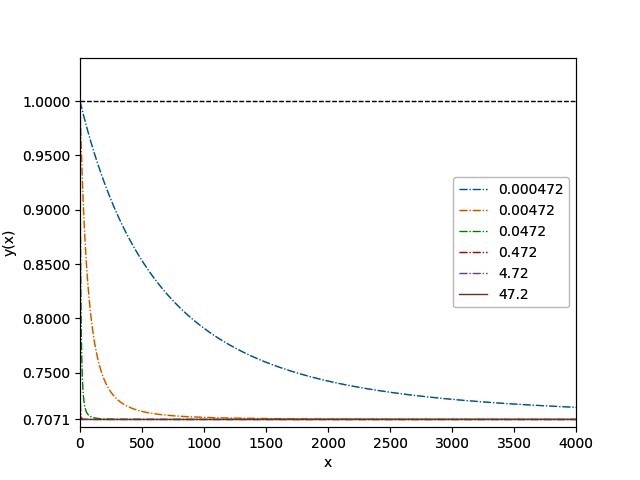}
\caption{All the normalized distributions $y(x)$ with minimal length modification asymptotically goes to $1/\sqrt{2}$ for $\abs{\mathbf{p}}\to \infty$. For large $a$, we can have $\abs{\mathbf{E}_0}_x\approx 1/\sqrt{2b}\left[\Theta(x-x_c)+\sqrt{2}~\Theta(x_c-x)\right]$, where $x_c\approx 0^+$ corresponds to the value of $\abs{\mathbf{p}}$ for which, the distribution inflects sharply.}
\end{minipage}
\end{figure}
The critical field strength can be investigated while considering the limit--- $\displaystyle{\lim_{r\rightarrow 0}\abs{\mathbf{D}^{(a^2)}_0(\mathbf{r})}}$ where a consistent solution is obtained for:
\begin{equation}
\begin{aligned}
    1-b \abs{\mathbf{E_0}}^2+{ b a^2}\left({\Tilde{\omega}_{\mathbf{p}}^{(a)}}\right)^2 \abs{\mathbf{E_0}}^2&=0\\[4pt]
   \text{or,~~~~~~~} \abs{\mathbf{E_0}}^{(a^2)}_{\mathbf{p}}=\frac{1}{\sqrt{b}}\sqrt{\frac{1}{2-\frac{a^2\abs{\mathbf{p}}^2}{\hbar^2}\left(1-\sqrt{1+\frac{\hbar^2}{a^2\abs{\mathbf{p}}^2}}\right)^2}}~~.\\
\end{aligned}
\end{equation}
which takes a constant value $\displaystyle{\frac{1}{\sqrt{b}}}$ in the limit $a\to 0$ \cite{kruglov2015model} and $\displaystyle{\frac{1}{\sqrt{2b}}}$, when $\abs{\mathbf{p}}\to \infty$. It is to be noted that the above expression is obtained in the light of the leading order approximation to self-consistent equations (eqn.- \ref{equation-22}).

We further observe the variation of field strength at the origin with the strength of photon momentum and hence the energy spectrum from dispersion relation. Such variation can hence be a possible consequence of minimal length due to the coupling between $a$ and $\abs{\mathbf{p}}$. It is further expected that such variations would persist for higher order approximations as well. We can thus study the above variation as a distribution of field strength values in $\abs{\mathbf{p}}$- space by taking $\displaystyle{x\equiv\frac{\abs{\mathbf{p}}}{\hbar}}$ and $y\equiv \abs{\mathbf{E}}_{\mathbf{p}}^{(a^2)}\sqrt{b}$ ~~(see figs.- \textcolor{blue}{1,2}).\\

 Quite often, phenomenological analyses relies on fixing spectral bounds to tame divergences, which can be considered as a method of mere computational expediency. It can however, be asserted that such bounds are physical and are of cosmological origin \cite{delphenich2003nonlinear} or might be possibly related to the micro-structure of the background. Similar type of bound is associated with the limitation of perturbative approaches in QED ~(\textit{e.g.- Schwinger} limit in perturbative QED) which demands non-linear modification to the \textit{Maxwellian} field theory \cite{buchanan2006past}. 


\subsection{Vacuum birefringence in minimal length formulation}
In the context of effective action approach, birefringence has been thoroughly studied in the presence of external magnetic field from spinor QED \cite{dittrich1998vacuum} by considering the one-loop effective action.
However, we consider an electromagnetic field characterized by the disturbance ($\mathbf{\mathcal{E}}, \mathbf{\mathcal{B}}$), which propagate in a preferred direction (along z-axis) following the dispersion relations obtained from linear electrodynamics, in the absence of minimal length. In the presence of a constant, uniformly strong magnetic perturbation, $\mathbf{B}_{ext}=(0, B, 0)$ with $\abs{\mathbf{B}_{ext}}>>\abs{\mathbf{\mathcal{E}}}, \abs{\mathbf{\mathcal{B}}}$, non linear effects are expected to override and the resultant field characterized by ($\mathbf{\mathcal{E}}, \mathbf{\mathcal{B}}+\mathbf{B}_{ext}$) should agree with the non-linear corrections as motivated in the literature \cite{brezin1971polarization, shibata2020intrinsic, denisov2017nonperturbative} along with the minimal length modifications. 
\\[2pt]
Following the corrections motivated from peturbative quantum gravity, one can analyze different polarizations of the field disturbance by similar arguments as in \cite{kruglov2015nonlinear}. This leads to the following expressions for magnetic permeability and electric permittivity (from eqn.- \ref{equation-28}) [Here, we take $c=1,  \mu_0=1 ~\&~ \epsilon_0=1$]:
\begin{equation*}
    \epsilon_{\gamma~~~ij}^{(a^2)}=\epsilon^{(a^2)}~\delta_{ij}+\Gamma_{\gamma}^{(a^2)}~\delta_{i2} \delta_{j2}~B^2~~
    \text{~~and~~~~~~}
    \mu_{\gamma~~~ij}^{(a^2)}=\displaystyle{\frac{1}{\epsilon^{(a^2)}}}~\delta_{ij}~~;
\end{equation*}
where,
\begin{equation*}
     \epsilon^{(a^2)}=\left[\Bigg(1-a^2\left({\Tilde{\omega}_{\mathbf{p}}^{(a)}}\right)^2\Bigg)+ \frac{\alpha \Bigg(1-a^2 \left({\Tilde{\omega}_{\mathbf{p}}^{(a)}}\right)^2\Bigg)}{\Bigl(1+{b}\abs{\mathbf{B}_{ext}}^2-{ba^2}\left({\Tilde{\omega}_{\mathbf{p}}^{(a)}}\right)^2 \abs{\mathbf{B}_{ext}}^2\Bigr)^2}\right]~~.
\end{equation*}

For parallel polarization in the direction of the external field, we get:
\begin{equation}
    n_{\parallel}=\frac{1}{v_{\parallel}}= \sqrt{\frac{\epsilon^{(a^2)}_{\gamma~~~22}}{\epsilon^{(a^2)}}}~~.
\end{equation}
Likewise, for transverse polarization, the refractive index is unchanged and hence we obtain \cite{kruglov2015nonlinear}:
\begin{equation*}
    n_{\perp}=1
\end{equation*}
and the phase velocity remains equal to the speed of light i.e. $v_{\perp}=c=1$ (by our choice).
Such unequal velocity in mutually perpendicular, polarization directions is captured in the following expression for $\Delta n \equiv n_{\parallel}-n_{\perp}$:

\begin{equation}\label{equation-32}
    \Delta n^{(a^2)}_{\mathbf{p}}=\sqrt{1+\frac{{\gamma}~K_2 \abs{\mathbf{B}_{ext}}^2\left(1+{b}K_1 \abs{\mathbf{B}_{ext}}^2\right)^2}{\alpha K_1+\left(1+{b}K_1 \abs{\mathbf{B}_{ext}}^2\right)^2}}-1~~
\end{equation}
with,
\begin{equation*}
    \begin{aligned}
    K_1&\equiv 1-a^2\left({\Tilde{\omega}_{\mathbf{p}}^{(a)}}\right)^2 \text{~~~and}\\
    K_2&\equiv 1-2a^2\left({\Tilde{\omega}_{\mathbf{p}}^{(a)}}\right)^2 ~~.
    \end{aligned}
\end{equation*}
It is to be noted that $\Delta n$ depends on photon momentum due to minimal length modifications. Moreover, we observe that the expected value of $\Delta n$ is larger than the value predicted in the absence of minimal length (see figs.- \textcolor{blue}{3,4}). 

\begin{figure}[ht]
\centering 
\begin{minipage}[t]{6cm}
\centering
\includegraphics[scale=0.45]{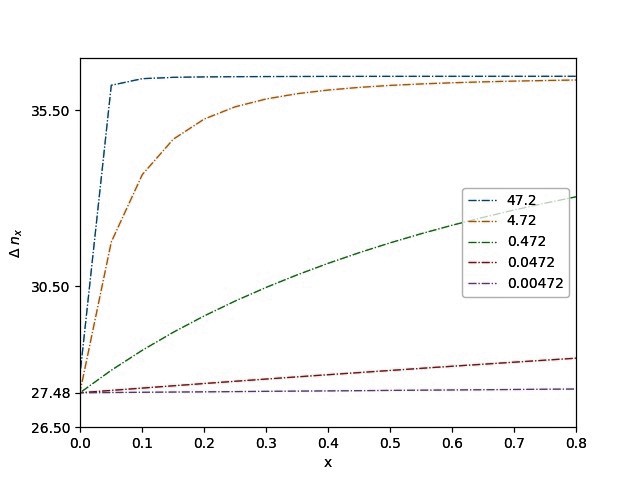}
\caption{The above schematic (scaled appropriately) depicts the behavior of $\Delta n$ as a function of $x\equiv \abs{\mathbf{p}}/\hbar$. The functional behavior is independent of the choice of parameters: $\gamma$, $b$ and $\alpha$ however we observe variation for different values of characteristic length parameter $a \times 10^{-17} cm$. For the sake of convenient graphical demonstration, we choose: $\gamma=0.00001 \times 10^{-15}~T^{-2}$, $\abs{\mathbf{B}_{ext}}=0.9\times 10^{10}T$ so that, $\gamma\abs{\mathbf{B}_{ext}}^2\sim \mathcal{O}(1)$. Here, we observe that, for $\abs{\mathbf{p}}\to 0$ approximation, $\Delta n$ agrees with predictions consistent with classical background (with well-defined localization).}
\end{minipage}
\hspace{2.1cm}
\begin{minipage}[t]{6cm}
\includegraphics[scale=0.45]{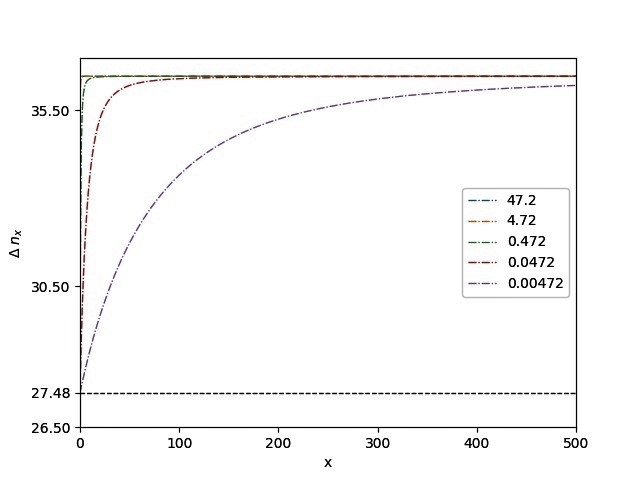}
\caption{The above schematic depicts the large $\abs{\mathbf{p}}$ behavior of $\Delta n$. Here, the black, horizontal dashed line depicts the value of $\Delta n$ when minimal length is ignored. In the presence of $a$, the functional behavior is that of an extremely slowly-growing function for $\abs{\mathbf{p}}\to \infty$ (although, it is suggested that $\abs{\mathbf{p}}<p_{max}$ ,where $p_{max}$ is the upper momentum cut-off, from higher order GUP theories \cite{pedram2012higher, nozari2012natural}). Thus, such functional behavior indicates--- $\Delta n_{\mathbf{p}^{(a^2)}}>\Delta n_{(a^2=0)}$ for all values of the other three parameters and $\abs{\mathbf{B}_{ext}}$.}
\end{minipage}
\label{Fig: 1}
\end{figure}

As a result, the parallel component of phase velocity of the disturbance of EM field (along the applied magnetic field), suffers a resistive force which can be modelled as: $f_{\parallel}^{(a^2)}\sim \abs{{\mathbf{p}}}^{\delta}\abs{{\mathbf{B}_{ext}}}^{\lambda}$ (for some $\lambda>0$ and $0<\delta<1$, which must come from a fundamental theoretical description), possibly from the quantized property of the background.


\section{Discussion}

A possible modification to \textit{Kruglov}'s non-linear electrodynamics is suggested in the light of GUP. By, considering perturbative quantum-gravity effects in leading order, we make modifications to the model, whereupon we obtain the zero-point field strength. The behavior of zero-point field strength, depicts a variation with photon momentum. Lastly, by studying vacuum birefringence in the presence of external, constant and uniform magnetic field, we conclude a momentum dependency in the behavior of $\Delta n$. The minimal length modifications also suggest, that the velocity along parallel direction of polarization is lower than the usual value i.e. $\displaystyle{v_{\parallel^{(a^2)}}<v_{\parallel}}$. This is due to the presence of an unavoidable coupling between the momentum of the field quanta (photon) and the minimum permissible length associated with the quanta of the background.\\[2pt]
Such theoretical implications suggest the presence of a frictional mechanism (slowing down of velocity in the direction of applied field), possibly originating from the discreteness of the background. Existence of similar quantum-gravitational friction, affecting photons moving in space-time foam has also been discussed by \textit{J. Ellis} et al. \cite{ellis1992string}.


\section*{Acknowledgments}
We sincerely acknowledge \textit{Vikramaditya Mondal} for enlightening us with useful discussions and insightful comments.
This research is funded by the University Grants Commission (UGC), Government of India.


\bibliographystyle{unsrt}
\bibliography{mybib}

\begin{thebibliography}{10}

\bibitem{kiefer2013conceptual}
Claus Kiefer.
\newblock Conceptual problems in quantum gravity and quantum cosmology.
\newblock {\em International Scholarly Research Notices}, 2013, 2013.

\bibitem{ziaeepour2022comparing}
Houri Ziaeepour.
\newblock Comparing quantum gravity models: String theory, loop quantum
  gravity, and entanglement gravity versus su ($\infty$)-qgr.
\newblock {\em Symmetry}, 14(1):58, 2022.

\bibitem{hossenfelder2006note}
S~Hossenfelder.
\newblock A note on theories with a minimal length.
\newblock {\em Classical and Quantum Gravity}, 23(5):1815, 2006.

\bibitem{ashtekar2004background}
Abhay Ashtekar and Jerzy Lewandowski.
\newblock Background independent quantum gravity: a status report.
\newblock {\em Classical and Quantum Gravity}, 21(15):R53, 2004.

\bibitem{douglas2001noncommutative}
Michael~R Douglas and Nikita~A Nekrasov.
\newblock Noncommutative field theory.
\newblock {\em Reviews of Modern Physics}, 73(4):977, 2001.

\bibitem{girelli2005deformed}
Florian Girelli, Etera~R Livine, and Daniele Oriti.
\newblock Deformed special relativity as an effective flat limit of quantum
  gravity.
\newblock {\em Nuclear Physics B}, 708(1-3):411--433, 2005.

\bibitem{gross1988string}
David~J Gross and Paul~F Mende.
\newblock String theory beyond the planck scale.
\newblock {\em Nuclear Physics B}, 303(3):407--454, 1988.

\bibitem{thiemann2003lectures}
Thomas Thiemann.
\newblock Lectures on loop quantum gravity.
\newblock In {\em Quantum gravity}, pages 41--135. Springer, 2003.

\bibitem{yoneya1989interpretation}
Tamiaki Yoneya.
\newblock On the interpretation of minimal length in string theories.
\newblock {\em Modern Physics Letters A}, 4(16):1587--1595, 1989.

\bibitem{perez2003spin}
Alejandro Perez.
\newblock Spin foam models for quantum gravity.
\newblock {\em Classical and Quantum Gravity}, 20(6):R43, 2003.

\bibitem{konishi1990minimum}
Kenichi Konishi, Giampiero Paffuti, and Paolo Provero.
\newblock Minimum physical length and the generalized uncertainty principle in
  string theory.
\newblock {\em Physics Letters B}, 234(3):276--284, 1990.

\bibitem{veneziano1986stringy}
Gabriele Veneziano.
\newblock A stringy nature needs just two constants.
\newblock {\em EPL (Europhysics Letters)}, 2(3):199, 1986.

\bibitem{bosso2022minimal}
Pasquale Bosso, Luciano Petruzziello, and Fabian Wagner.
\newblock The minimal length is physical.
\newblock {\em Physics Letters B}, 834:137415, 2022.

\bibitem{kempf1995hilbert}
Achim Kempf, Gianpiero Mangano, and Robert~B Mann.
\newblock Hilbert space representation of the minimal length uncertainty
  relation.
\newblock {\em Physical Review D}, 52(2):1108, 1995.

\bibitem{quesne2006lorentz}
Christiane Quesne and VM~Tkachuk.
\newblock Lorentz-covariant deformed algebra with minimal length.
\newblock {\em Czechoslovak Journal of Physics}, 56(10):1269--1274, 2006.

\bibitem{moayedi2013formulation}
SK~Moayedi, MR~Setare, and B~Khosropour.
\newblock Formulation of electrodynamics with an external source in the
  presence of a minimal measurable length.
\newblock {\em Advances in High Energy Physics}, 2013, 2013.

\bibitem{heisenberg2006consequences}
W~Heisenberg and H~Euler.
\newblock Consequences of dirac theory of the positron.
\newblock {\em arXiv preprint physics/0605038}, 2006.

\bibitem{kruglov2015nonlinear}
SI~Kruglov.
\newblock Nonlinear electrodynamics with birefringence.
\newblock {\em Physics Letters A}, 379(7):623--625, 2015.

\bibitem{Hattori_2013}
Koichi Hattori and Kazunori Itakura.
\newblock Vacuum birefringence in strong magnetic fields: (i) photon
  polarization tensor with all the landau levels.
\newblock {\em Annals of Physics}, 330:23--54, mar 2013.

\bibitem{mignani2017evidence}
Roberto~P Mignani, Vincenzo Testa, D~Gonzalez Caniulef, Roberto Taverna,
  Roberto Turolla, Silvia Zane, Kinwah Wu, and G~Lo Curto.
\newblock Evidence of vacuum birefringence from the polarisation of the optical
  emission from an isolated neutron star.
\newblock {\em arXiv preprint arXiv:1710.08709}, 2017.

\bibitem{mondal2020duality}
Vikramaditya Mondal.
\newblock Duality principle of the zero-point length of spacetime and
  generalized uncertainty principle.
\newblock {\em Europhysics Letters}, 132(1):10005, 2020.

\bibitem{frenkel1999self}
Josif Frenkel and RB~Santos.
\newblock The self-force of a charged particle in classical electrodynamics
  with a cutoff.
\newblock {\em International Journal of Modern Physics B}, 13(03):315--324,
  1999.

\bibitem{accioly2010limits}
Antonio Accioly and Eslley Scatena.
\newblock Limits on the coupling constant of higher-derivative
  electromagnetism.
\newblock {\em Modern Physics Letters A}, 25(04):269--276, 2010.

\bibitem{podolsky1942generalized}
Boris Podolsky.
\newblock A generalized electrodynamics part i—non-quantum.
\newblock {\em Physical Review}, 62(1-2):68, 1942.

\bibitem{denisov2017nonperturbative}
VI~Denisov, EE~Dolgaya, and VA~Sokolov.
\newblock Nonperturbative qed vacuum birefringence.
\newblock {\em Journal of High Energy Physics}, 2017(5):1--13, 2017.

\bibitem{lemos1999born}
Jos{\'e}~PS Lemos and Richard Kerner.
\newblock The born-infeld electromagnetism in kaluza-klein theory.
\newblock {\em arXiv preprint hep-th/9907187}, 1999.

\bibitem{kruglov2015model}
SI~Kruglov.
\newblock A model of nonlinear electrodynamics.
\newblock {\em Annals of Physics}, 353:299--306, 2015.

\bibitem{delphenich2003nonlinear}
David Delphenich.
\newblock Nonlinear electrodynamics and qed.
\newblock {\em arXiv preprint hep-th/0309108}, 2003.

\bibitem{buchanan2006past}
Mark Buchanan.
\newblock Past the schwinger limit.
\newblock {\em Nature Physics}, 2(11):721--721, 2006.

\bibitem{dittrich1998vacuum}
Walter Dittrich and Holger Gies.
\newblock Vacuum birefringence in strong magnetic fields.
\newblock {\em arXiv preprint hep-ph/9806417}, 1998.

\bibitem{brezin1971polarization}
E~Brezin and C~Itzykson.
\newblock Polarization phenomena in vacuum nonlinear electrodynamics.
\newblock {\em Physical Review D}, 3(2):618, 1971.

\bibitem{shibata2020intrinsic}
Kazunori Shibata.
\newblock Intrinsic resonant enhancement of light by nonlinear vacuum.
\newblock {\em The European Physical Journal D}, 74(10):1--6, 2020.

\bibitem{pedram2012higher}
Pouria Pedram.
\newblock A higher order gup with minimal length uncertainty and maximal
  momentum ii: Applications.
\newblock {\em Physics Letters B}, 718(2):638--645, 2012.

\bibitem{nozari2012natural}
Kourosh Nozari and Sara Saghafi.
\newblock Natural cutoffs and quantum tunneling from black hole horizon.
\newblock {\em Journal of High Energy Physics}, 2012(11):1--18, 2012.

\bibitem{ellis1992string}
John Ellis, NE~Mavromatos, and Dimitri~V Nanopoulos.
\newblock String theory modifies quantum mechanics.
\newblock {\em Physics Letters B}, 293(1-2):37--48, 1992.

\end{thebibliography}

\end{document}